\documentclass[onecolumn,showpacs,preprintnumbers,amsmath,amssymb]{revtex4}

\usepackage{graphicx}
\usepackage{dcolumn}
\usepackage{bm}

\begin{document}

\preprint{APS/123-QED}

\title{$ \dfrac{1}{c^2} $ Correction to Thermodynamics}

\author{Jose A. Magpantay}
\email{jose.magpantay11@gmail.com}
\affiliation{Quezon City 1101, Philippines\\}

\date{\today}

\begin{abstract}
I work out the general expressions for the first relativistic correction of order $ \dfrac{1}{c^2} $ to thermodynamics. The starting point is the relativistic Hamiltonian that I derived in a previous paper, which I expanded to powers of $ \dfrac{1}{c^2} $ to derive a local (in time) Hamiltonian. Limiting to the first relativistic correction, I worked out in general how the relativistic corrections to thermodynamics arise. I then applied the formalism to the problem of N particles with harmonic oscillator interaction in 3D to see the explicit expressions for relativistic corrections. 
\end{abstract}

\pacs{Valid PACS appear here}
\maketitle

\section{\label{sec:level1}Introduction}
Laboratory gas systems at room temperature have non-relativistic velocities, typically of the order of $ 10^3 \frac{m}{s} $, which makes standard non-relativistic thermodynamics valid. However, there are systems, example, fusion reactors, where the velocities are relativistic (velocities of the order of $ 10^6 \frac{m}{s} $) but not by much, of the order of $ 1 \% $ of the speed of light. Relativistic corrections are small but still worthwhile to determine. This regime is the focus of this paper, to compute the relativistic correction to thermodynamic systems. Stellar systems, where ions are relativistic, are also in strong gravitational field so the method of general relativistic statistical mechanics is more apt.  For elementary particles approaching the speed of light, quantum effects are also significant and the method of quantum field theory, rather than particle-particle dynamics, is more relevant. 

The starting point of equilibrium thermodynamics is the partition function, which necessitates knowing the Hamiltonian of the system. When I wrote the paper \cite{Magpantay} that gave a rather simple relativistic Hamiltonian for a non-relativistic Hamiltonian  
\begin{equation}\label{1}
H_{nr} = \sum_{a=1}^{N} \frac{1}{2m} \vec{p_{a}} \cdot \vec{p_{a}} + \sum_{b<c}^{N} V(\vert\vec{x}^{b} - \vec{x}^{c}\vert),
\end{equation}
I was not aware of the works done on extending to relativistic dynamics. I was only familiar with the works that stated that there is no Hamiltonian for interacting particles that will satisfy the Poincare group \cite{Currie} \cite{Cannon} \cite{Leutwyler}  until I chanced upon a review article on relativistic brownian motion \cite{Dunkel}, which gave a short summary of attempts on the search for a relativistic many particle dynamics. There seems to be two approaches, one is by using Dirac's constrained formalism to derive a Hamiltonian that satisfies the Poincare algebra as represented by the following papers \cite{Komar} \cite{Goldberg} \cite{Kihlberg} \cite{Samuel}. It seems the Poincare algebra is satisfied in this approach but the non-local in time interaction is not transparent. Also, the Hamitonian is not unique as it depends on the additional constraints one has to impose. 

The other is to take into account the non-local in time nature of relativistic dynamics as presented in the works of \cite{Woodcock} \cite{Marnelius}, with the last paper claiming that the non-local in time formulation leads to translation, rotation and boost generators that satisfy the Poincare algebra. This is where I differ because in my simple attempt to derive a many particle relativistic dynamics by making use of a scalar field and eventually integrating it out, the generators in the effective dynamics are conserved but do not satisfy the Poincare algebra because of the non-local in time dynamics. Still, the derived relativistic Hamiltonian nicely gives the non-relativistic two body interactions in the limit $ c \rightarrow \infty $. Thus, although my previous paper has not settled the Poincare algebra issue raised originally, the relativistic Hamiltonian I derived looks like a viable starting point for a relativistic thermodynamics. This is what I will do in this paper. And since an exact evaluation of the partition function is not possible, I will do an expansion in $ \frac{1}{c^{2}} $ corrections to the non-relativistic thermodynamics. 

But first, I give a summary of the relativistic Hamiltonian I presented previously. This is given by
\begin{subequations}\label{2}
\begin{gather}
H_{r} = \sum_{a=1}^{N} c(p_{a}^{2} + m^{2}c^{2})^{\frac{1}{2}} + \int dt' \sum_{b<c}^{N} G_{4}(\vec{x}^{b}(t)-\vec{x}^{c}(t'); t-t'), \label{first} \\
 G_{4}(\vec{x}^{b}(t)-\vec{x}^{c}(t'); t-t') = \int d^{3}k dk_{0} \dfrac{ \exp {[ik_{0}(t - t') - i \vec{k}\cdot(\vec{x}^{b}(t') - \vec{x}^{a}(t))}]}{n^{2}(\vert\vec{k}\vert^{2} - \dfrac{k_{0}^{2}}{c^{2}})}, \label{second} \\
 n^{2}(\vert\vec{k}\vert^{2}) =  (2\pi)^{3}\left[ \int d^{3}x V(\vert\vec{x}\vert) \exp {i\vec{k}\cdot\vec{x}} \right]^{-1},
 \end{gather}
 \end{subequations}
 where V is the non-relativistic two-body potential given in equation (1). Note that the argument of $ n^{2} $ solved in equation (2c) was changed from $ \vert\vec{k}\vert^{2} $ to $ \vert\vec{k}\vert^{2} - \dfrac{k_{0}^{2}}{c^{2}} $ in equation (2b). 
 This simple prescription gives the relativistic version of equation (1). 
 
 Even if we find a closed form of $ G_{4} $, the non-locality in time in $ H_{r} $ will make its use in the evaluation of the partition function rather too complicated. A way out is to expand in powers of $ \frac{1}{c^{2}} $, which I will show how to do in the next section.
 
\section{\label{sec:level2}The Expansion}

The starting point is the expansion of $ n^{2} $ as given by
\begin{equation}\label{3}
\begin{split}
n^{2}(\vert\vec{k}\vert^{2} - \dfrac{k_{0}^{2}}{c^{2}})& = n^{2}(\vert\vec{k}\vert^{2}) + (-\dfrac{k_{0}^{2}}{c^{2}}) \dfrac{\partial n^{2}(\vert\vec{k}\vert^{2})}{\partial \vert\vec{k}\vert^{2}}\\ 
& \quad + \frac{1}{2} (-\dfrac{k_{0}^{2}}{c^{2}})^{2} \dfrac{\partial^{2} n^{2}(\vert\vec{k}\vert^{2})}{\partial \vert\vec{k}\vert^{2}\partial \vert\vec{k}\vert^{2}} + ....
\end{split}
\end{equation}
Substituting this in equation (2b) gives 
\begin{equation}\label{4}
\begin{split}
G_{4}(\vec{x}^{b}(t)-\vec{x}^{c}(t'); t-t')& = \delta(t-t') V(\vert\vec{x}^{b}(t)-\vec{x}^{c}(t')\vert)\\ 
& \quad + \frac{1}{c^{2} }\int d^{3}k dk_{0} k_{0}^{2} \exp {[ik_{0}(t-t')]} \dfrac{1}{[n^{2}(\vert\vec{k}\vert^{2})]^{2}} \dfrac{\partial n^{2}(\vert\vec{k}\vert^{2})}{\partial \vert\vec{k}\vert^{2}} \exp {[-i\vec{k}\cdot(\vec{x}^{b}(t) - \vec{x}^{c}(t'))]} + ...
\end{split}
\end{equation}
where the ellipsis represent higher orders in $ \frac{1}{c^{2}} $. Substituting this expression in equation (2a) gives
\begin{subequations}\label{5}
\begin{gather}
H_{r} = \sum_{a=1}^{N} c(p_{a}^{2} + m^{2}c^{2})^{\frac{1}{2}} + \sum_{b<c}^{N} V(\vert\vec{x}^{b} - \vec{x}^{c}\vert) - \frac{1}{c^{2}} \sum_{b<c}^{N} \left[ i \ddot{x}^{c}_{j}(t) L_{j}(\vec{x}^{b} - \vec{x}^{c}) + \dot{x}^{c}_{j}\dot{x}^{c}_{k} Q_{jk}(\vec{x}^{b} - \vec{x}^{c}) \right], \label{first}\\
L_{j}(\vec{x}^{b} - \vec{x}^{c}) = \int d^{3}k k_{j}\dfrac{1}{[n^{2}]^{2}}\dfrac{\partial n^{2}(\vert\vec{k}\vert^{2})}{\partial \vert\vec{k}\vert^{2}} \exp {[-i\vec{k}\cdot(\vec{x}^{b}(t) - \vec{x}^{c}(t))]}, \label{second}\\
Q_{jk}(\vec{x}^{b} - \vec{x}^{c}) = \int d^{3}k k_{j}k_{k} \dfrac{1}{[n^{2}]^{2}}\dfrac{\partial n^{2}(\vert\vec{k}\vert^{2})}{\partial \vert\vec{k}\vert^{2}} \exp {[-i\vec{k}\cdot(\vec{x}^{b}(t) - \vec{x}^{c}(t))]}.
\end{gather}
\end{subequations}
Equation (5) clearly shows that as $ c \rightarrow \infty $, we get the non-relativistic Hamiltonian given by equation(1). 

Note, equation (5) is still expressed in terms of velocities, which must be given in terms of momenta. It even has an acceleration term, which we now argue as negligible compared to the velocity terms of the same order in $ \frac{1}{c^{2}} $. Firstly, the acceleration term $ \dfrac{\ddot{x}}{c^{2}} $ is definitely much smaller than the velocity term $ [\dfrac{\dot{x}}{c}]^{2} $. Second, the integral factor that goes with the acceleration term has an integrand with one power of $ \vec{k} $ less than the velocity term. Thus, we can neglect the third term of equation (5a).

The relativistic velocity in terms of momentum is given by
\begin{equation}\label{6}
\dot{x}^{b}_{i} = \frac{1}{m} \dfrac{{p}^{b}_{i}}{(1 + \dfrac{p_{b}^{2}}{m^{2}c^{2}})^{\frac{1}{2}}}
\end{equation}.
Substituting this in equation (5a) and consistently expanding all terms to $ \frac{1}{c^{2}} $, we find 
\begin{subequations}\label{7}
\begin{gather}
H_{r} = H_{nr} + \frac{1}{c^{2}} \Delta H_{1} + ...., \label{first}\\
\Delta H_{1} = -\left[ \frac{1}{8m^{3}}\sum_{a=1}^{N} (p_{a})^{4} +\frac{1}{m^{2}}\sum_{b<c}^{N} p^{c}_{j}p^{c}_{k} Q_{jk}(\vec{x}^{b} - \vec{x}^{c})\right], 
\end{gather}
\end{subequations}
where the ellipsis in equation (7a) represent terms that are of order $ \frac{1}{c^{4}} $ or higher. 

Thermodynamics begins with the evaluation of the partition function given by
\begin{equation}\label{8}
Z_{r} = \frac{1}{N!} \int \prod_{a=1}^{N} d\vec{x}^{a} d\vec{p}^{a} \exp {(-\beta H_{r})},
\end{equation}
which upon using equation (7) gives
\begin{subequations}\label{9}
\begin{gather}
Z_{r} = Z_{nr} + \frac{1}{c^{2}} \Delta Z, \label{first}\\
Z_{nr} = \frac{1}{N!} (\frac{1}{h^{3}})^{N} \int \prod_{a=1}^{N} d\vec{x}^{a} d\vec{p}^{a} \exp {(-\beta H_{nr})},\label{second}\\
\Delta Z = \beta \frac{1}{N!} (\frac{1}{h^{3}})^{N} \int \prod_{a=1}^{N} d\vec{x}^{a} d\vec{p}^{a} \Delta H_{1} \exp {(-\beta H_{nr})}.
\end{gather}
\end{subequations}
Equation (9) clearly gives the $ \frac{1}{c^{2}} $ corrections to thermodynamics, which we will explicitly compute for a specific two-body potential. 

\section{\label{sec:level3}Non-Relativistic Harmonic Oscillator Potential in 3D}
Consider N identical particles of mass m, confined in a cube of volume $ V = L^{3} $, and at temperature T. Assume that the particles interact with the harmonic potential 
\begin{equation}\label{10}
V(\vert\vec{x}^{b}-\vec{x}^{c}\vert) = \frac{\lambda}{2} \vert\vec{x}^{b}-\vec{x}^{c}\vert^{2},
\end{equation}
where the strength of interaction $ \lambda $ is the same for all two-particle interactions. 

If the gas is non-relativistic, the partition function, as given by equation (9b), is
\begin{equation}\label{11}
Z_{nr} = \frac{1}{N!} \int \prod_{a=1}^{N} d\vec{x}^{a} d\vec{p}^{a} \exp {(-\beta\left[ \sum_{a=1}^{N} \frac{1}{2m} \vec{p}_{a}^{2} + \frac{\lambda}{2} \sum_{b<c}^{N} \vert\vec{x}^{b}-\vec{x}^{c}\vert^{2}\right] )}.
\end{equation}
The momentum integrations are trivial, they give $ (4\pi)^{N}(\dfrac{2\pi m}{\beta})^{\frac{3N}{2}} $ while the coordinate integrals are a bit more involved. The container is defined by having $ x_{i} = [-\frac {L}{2}, \frac{L}{2}]$, for $i = 1, 2,3 $. The integration involving $ \vec{x}_{1} $ is carried out by making a change of variables $ \vec{x'}_{1} = \vec{x}_{1} - \frac{1}{N-1}\sum_{c=2}^{N} \vec{x}_{c} $. The $ \vec{x'}_{1} $ integration gives $ (\dfrac{\sqrt{\pi}}{2\alpha_{1}})^{3} \Phi^{3}(\alpha_{1}V^{\frac{1}{3}}) $, where $ \Phi(x) $ is the error function defined by the integral
\begin{equation}\label{12}
\Phi(x) = \frac{2}{\sqrt{\pi}} \int_{0}^{x} dt \exp {(-t^{2})},
\end{equation}
where $ \alpha_{1} = \frac{1}{2\sqrt{2}}[\beta \lambda (N-1)]^{\frac{1}{2}} $. 

Next, integrate $ \vec{x}_{2} $. Taking into account the additional terms when we defined $ \vec{x'}_{1} $ and the other terms involving $ \vec{x}_{2} $ coming from the two-body potential terms, the change of variable $ \vec{x'}_{2} = \vec{x}_{2} - \frac{1}{N-2} \sum_{c=3}^{N} \vec{x}_{c} $. Integrating $ \vec{x'}_{2} $ gives the contribution $ (\dfrac{\sqrt{\pi}}{2\alpha_{2}})^{3} \Phi^{3}(\alpha_{2}V^{\frac{1}{3}}) $ where $ \alpha_{2} = \frac{1}{2\sqrt{2}}[\beta \lambda \dfrac{N(N-2)}{(N-1)}]^{\frac{1}{2}} $. 

Continuing the process to step j for $ j =3, 4, 5 $, there is a noticeable pattern, that for an arbitrary j, define $ \vec{x'}_{j} = \vec{x}_{j} - \frac{1}{N-j} \sum_{c=j+1}^{N} \vec{x}_{c} $. Integrating $ \vec{x'}_{j} $ gives the contribution $ (\dfrac{\sqrt{\pi}}{2\alpha_{j}})^{3} \Phi^{3}(\alpha_{j}V^{\frac{1}{3}}) $, where $ \alpha_{j} = \frac{1}{2\sqrt{2}}[\beta \lambda \dfrac{N(N-j)}{(N-j+1)}]^{\frac{1}{2}} $. To prove that the formulas at level j are valid, I make use of induction. I assume that the jth step formulas are valid and derive the $ (j+1)^{st} $ formulas and show that these are precisely given by the $ j^{th} $ iteration formulas with j changed to $ j+1 $. 

Then we follow the process until the $ N-1 $ term to give the last change of variables $ \vec{x'}_{N-1} = \vec{x}_{N-1} -\vec{x}_{N} $. Integrate this term to get the contribution $ (\dfrac{\sqrt{\pi}}{2\alpha_{N-1}})^{3} \Phi^{3}(\alpha_{N-1}V^{\frac{1}{3}}) $, where $ \alpha_{N-1} = \frac{1}{2\sqrt{2}}[\beta \lambda \frac{N}{2}]^{\frac{1}{2}} $. Finally, the last integration over $ \vec{x}_{N} $ is done with result similar to that of $ \vec{x}_{1} $. 

The simplification of all the terms follow from the fact that we have typically $ N \propto 10^{23} $, thus all the $ \alpha_{j} = \frac{1}{2\sqrt{2}}[\beta \lambda N]^{\frac{1}{2}} $, for $ j = 1,... N $ are for all intents and purposes the same as $ \alpha = \frac{1}{2\sqrt{2}}[\beta \lambda N]^{\frac{1}{2}} $. The resulting non-relativistic partition function is
\begin{equation}\label{13}
Z_{nr} = \frac{1}{N!}(\frac{1}{h^{3}})^{N}(4\pi)^{N} \left( \dfrac{2\pi m}{\beta}\right)^{\frac{3}{2}N} \left( \dfrac{\sqrt{\pi}}{2\alpha} \right)^{3N} \Phi^{3N}(\alpha V^{\frac{1}{3}}).
\end{equation}

This is one of the main results of the paper, the partition function for a thermodynamic system of non-relativistic N particles in 3D with two-body harmonic interactions in a closed form. However using equation (12), the partition function is not very useful. But there is a useful series expansion of the error function, the Burman series \cite{Wikipedia} given by
\begin{equation}\label{14}
\begin{split}
\Phi(x)& = \frac{2}{\sqrt{\pi}} sgn(x) \sqrt{1 - \exp {(-x^{2})}}\left[ \frac{\pi}{2} + \sum_{k=1}^{\infty} c_{k}\exp {(-k x^{2})} \right]\\
				& = \frac{2}{\sqrt{\pi}} sgn(x) \sqrt{1 - \exp {(-x^{2})}}\left[ \frac{\pi}{2} + \frac{31}{100} \exp {(-x^{2})} - \frac{341}{8000} \exp {(-2x^{2})} + ....\right],
\end{split}
\end{equation}
where $ c_{k} $ are numbers, decreasing as k increases, and $ x = \alpha V^{\frac{1}{3}} $. Given $ \beta = \frac{1}{kT} $ and $ k \propto 10^{-23} $, and for thermodynamic systems $ N \propto 10^{23} $, we find x is indeed very large and we can, to a very good approximation, truncate $ \Phi(x) $ to 
\begin{equation}\label{15}
\Phi \approx 1 -(\frac{1}{2} - \frac{62}{100\sqrt{\pi}}) \exp {(-\alpha^{2} V^{\frac{2}{3}})} + ...
\end{equation}
Equation (15) is then substituted in equation(13) to give the non-relativistic partition function
\begin{equation}\label{16}
Z_{nr} \approx \frac{1}{N!}(\frac{1}{h^{3}})^{N}(4\pi)^{N} \left( \dfrac{2\pi m}{\beta}\right)^{\frac{3}{2}N}\left( \frac{2\pi}{\beta \lambda N} \right)^{\frac{3}{2}N} \left[ 1 - 3Na \exp {(-\frac{1}{8}\beta \lambda N V^{\frac{2}{3}})}\right],
\end{equation}
where $ a = \frac{1}{2} - \frac{62}{100\sqrt{\pi}} $. Although this partition function is not exact, the dropped terms are really small.  

Equation (16) is the starting point of the computation of the thermodynamic quantities, see for example \cite{Reif}, resulting in
\begin{subequations}\label{17}
\begin{gather}
E =  \frac{3N}{\beta} - \frac{3}{8} \lambda N^{2}aV^{\frac{2}{3}} \exp {(-\frac{1}{8}\beta \lambda N V^{\frac{2}{3}})}, 
	\label{first}\\
p = \frac{1}{4} N^{2}\lambda V^{-\frac{1}{3}} \exp {(-\frac{1}{8}\beta \lambda N V^{\frac{2}{3}})}, \label{second} \\
S = k\left[ \ln Z_{nr} +\beta E \right],
\end{gather}
\end{subequations}	
where the entropy S follows from equations (16) and (17a).

Note, the N non-relativistic free particles thermodynamics cannot be derived from above by taking $ \lambda \rightarrow 0 $. This is due to the fact that the above results follow from the truncated Burman expansion of the error function as given in equation (14), which is valid for large $ x = \frac{1}{2\sqrt{2}}(\beta \lambda N)^{\frac{1}{2} }V^{\frac{1}{3}} $. The free particle limit takes $ \lambda \rightarrow 0 $, which means taking x small and this necessitates another expansion for the error function, which is given by \cite{Gradshteyn}
\begin{equation}\label{18}
\begin{split}
\Phi(x)& = \frac{2}{\sqrt{\pi}} \sum_{k=1}^{\infty} (-1)^{k+1} \dfrac{x^{2k-1}}{(2k-1)(k-1)!}\\
			  &  \approx \frac{2}{\sqrt{\pi}} x + ...,
\end{split}
\end{equation}
where only the first term is explicitly written for very small x. Substituting this term in $ Z_{nr} $ given by equation (13), the result is simply
\begin{equation}\label{19}
Z_{nr} = \frac{1}{N!}(\frac{1}{h^{3}})^{N}(4\pi)^{N} \left( \dfrac{2\pi m}{\beta}\right)^{\frac{3}{2}N} V^{N},
\end{equation}
the known result for free particles from which follows the thermodynamic relations.
\section{\label{sec:level4}The $ \frac{1}{c^{2}} $ Relativistic Correction}
The discussions in Section II give the relativistic Hamiltonian given a non-relativistic system with 2 body potential. In particular, equation (7b) gives the first relativistic correction, the $ \frac{1}{c^{2}} $ correction to the non-relativistic Hamiltonian given by equation (1). I now apply this formalism to the system described in Section III. 

Using equation (2c), the relevant $ n^{2}(\vert\vec{k}\vert^{2}) $ is
\begin{equation}\label{20}
n^{2}(\vert\vec{k}\vert^{2}) = \dfrac{4\pi^{2}}{\lambda \Lambda} [\vert\vec{k}\vert^{2}]^{\frac{5}{2}},
\end{equation}
where $ \Lambda $ is a constant given by
\begin{equation}\label{21}
\Lambda = \int_{0}^{\infty} dr r^{3} \sin(r).
\end{equation}
As the discussions in the next sections will show, the relevant constant that appear in the relativistic correction is not $ \Lambda $ alone but  $ \Lambda $ multiplied by another constant $ \Xi $ and its is this product that will be evaluated in the appendix carefully. 

Important in equation (7b) are the $ Q_{jk}(\vec{x}^{b} - \vec{x}^{c}) $ given in equation(5c) and the result follows by substituting equation (20)
\begin{equation}\label{22}
Q_{jk}(\vec{x}^{b} - \vec{x}^{c}) = \frac{10}{\pi}\lambda(\Lambda \Xi)\left[ \vert\vec{x}^{b} - \vec{x}^{c}\vert^{2} \delta_{jk} + 2(x_{j}^{b} - x_{j}^{c})(x_{k}^{b} - x_{k}^{c}) \right] ,
\end{equation}
where $ \Lambda $ is given by equation(20) and $ \Xi $ is a constant given by 
\begin{equation}\label{23}
\Xi = \int_{0}^{\infty} dr \dfrac{\sin r}{r^{6}}.
\end{equation}
Equation (22) shows that the relativistic correction depends on the product of the two constants $ \Lambda \Xi $, which is evaluated in the Appendix A showing the value
\begin{equation}\label{24}
 \Lambda \Xi = -\frac{\pi}{40}.
 \end{equation}
 
Substituting in equations (7a) and (7b), the relativistically corrected Hamiltonian to $ H_{nr} $ given by equation(1) with the two-body potential given by equation(10) to $ \frac{1}{c^{2}} $ is
\begin{subequations}\label{25}
\begin{gather}
H_{r} = H_{nr} + \frac{1}{c^{2}} \Delta H_{1},\label{first}\\
\Delta H_{1} = -\frac{1}{8m^{3}} \sum_{a=1}^{N} (\vec{p}^{a}\cdot\vec{p}^{a})^{2} + \frac{\lambda}{4m^{2}}\sum_{b<c}^{N}\left[ (\vec{p}^{c}\cdot\vec{p}^{c})\vert \vec{x}^{b} - \vec{x}^{c} \vert^{2} + (\vec{p}^{c}\cdot(\vec{x}^{b} - \vec{x}^{c}))^{2} \right].
\end{gather}
\end{subequations}
Substituting equation (25b) in equation (9c) will solve for $ \Delta Z $ that should be added to $ Z_{nr} $ given by equations (13) and (15). The evaluation is tedious and long but straightforward giving 
\begin{equation}\label{26}
\Delta Z = \frac{9}{8m\beta} Z_{nr} -\frac{3\pi aN}{2m\beta}\left( \frac{2\pi}{\beta \lambda N}\right)^{\frac{3}{2}N-1} V^{\frac{2}{3}-N} \exp {(-\frac{1}{8}\beta} \lambda N V^{\frac{2}{3}} Z_{fnr}, 
\end{equation}
where $ Z_{nr} $ is the non-relativistic partition function for the N particles with harmonic oscillator two-body interaction given by equations (13) and (15) while $ Z_{fnr} $ is the partition function for N free non-relativistic particles given by
\begin{equation}\label{27}
Z_{fnr} =  \frac{1}{N!}(\frac{1}{h^{3}})^{N}(4\pi)^{N} \left( \dfrac{2\pi m}{\beta}\right)^{\frac{3}{2}N} V^{N}.
\end{equation}

From this, we can compute the $ \frac{1}{c^{2}} $ corrections to the thermodynamic quantities given in equations (17;a,b,c). 
\section{\label{sec:level5}The hybrid expansion}
Note that for a system of N free relativistic particles, the partition function is exactly solvable, i.e., 
\begin{equation}\label{28}
\begin{split}
Z_{fr}& = \frac{1}{N!}(\frac{1}{h^{3}})^{N} \int \prod_{a=1}^{N} d\vec{p}^{a} d\vec{x}^{a} \exp {(-\beta \sum_{a=1}^{N} c(\vec{p^{a}}^{2} + m^{2}c^{2})^{\frac{1}{2}})}\\
			&  =  Z_{rkin}V^N,
\end{split}
\end{equation}
where
\begin{equation}\label{29}	
Z_{rkin} = \frac{1}{N!}(\frac{1}{h^{3}})^{N} \left[ 4\pi \frac{m^{2}c}{\beta} K_{2}(mc^{2}\beta) \right]^{N},
\end{equation}
and $ K_{2} $ is a modified Bessel function of the second kind of order 2 \cite{Gradshteyn} \cite{Weisstein} and $ Z_{rkin} $ is the kinetic term of the free relativistic particle partition function. This suggests a possible more accurate expansion to order $ \frac{1}{c^{2}} $ of the partition function if the full relativistic kinetic term is accounted. This is what I call a hybrid expansion. Note, the qualifier possible because the correction, as the calculations in Appendix B show, may just involve higher powers of $ c^{2} $, which fortunately was shown to involve only higher powers of $ \frac{1}{c^{2}} $. The hybrid expansion can consistently make use of the full relativistic kinetic term and consistently expand the two-body interaction term in powers of $ \frac{1}{c^{2}} $.  

The hybrid expansion begins with the Hamiltonian given by equation (2a) with the two-body potential given by equation (10) but only expand to order $ \frac{1}{c^{2}} $ the relativistic potential term $ G_{4} $. The hybrid Hamiltonian is then given by
\begin{subequations}\label{30} 
\begin{gather}
H_{rhy} = H_{hy} + \frac{1}{c^{2}} \Delta H_{2}, \label{first}\\
H_{hy} = \sum_{a=1}^{N} c(\vec{p_{a}}^{2} + m^{2}c^{2})^{\frac{1}{2}} + \frac{\lambda}{2} \sum_{b<c}^{N} \vert\vec{x}^{b} - \vec{x}^{c}\vert^{2},\label{second}\\
\Delta H_{2} = +\frac{\lambda}{4 m^{2}} \sum_{b<c}^{N} p_{j}^{c}p_{k}^{c}\left[ \delta_{jk}\vert\vec{x}^{b} - \vec{x}^{c}\vert^{2} + 2(x_{j}^{b} - x_{j}^{c})(x_{k}^{b} -  x_{k}^{c}) \right] .
\end{gather}
\end{subequations}
The fact that the relativistic kinetic energy is added to the non-relativistic two body harmonic potential in equation (30b) shows why $ H_{hy} $ is a hybrid Hamiltonian. $ \Delta H_{2} $ given by equation (30c) differs from $ \Delta H_{1} $ given by equation (25b) because the kinetic term in equation (30b) is fully relativistic. From $  H_{hy} $ the hybrid partition function is easily computed using the spatial integration done in Section III and the momentum integration given in equation (29) resulting in
\begin{equation}\label{31}
Z_{hy} =  \frac{1}{N!}(\frac{1}{h^{3}})^{N} \left[ 4\pi \frac{m^{2}c}{\beta} K_{2}(mc^{2}\beta) \right]^{N}  \left( \frac{2\pi}{\beta \lambda N} \right)^{\frac{3}{2}N} \left[ 1 - 3Na \exp {(-\frac{1}{8}\beta \lambda N V^{\frac{2}{3}})}\right].
\end{equation}
The counterpart of equation (9) in this hybrid computation is
\begin{subequations}\label{32}
\begin{gather}
Z_{rhy} = Z_{hy} + \frac{1}{c^{2}} \Delta Z_{hy},\label{first}\\
\Delta Z_{hy} =\beta \frac{1}{N!}(\frac{1}{h^{3}})^{N} \int \prod_{a=1}^{N} d\vec{x}^{a} d\vec{p}^{a} \Delta H_{2} \exp {(- \beta H_{hy})}.
\end{gather}
\end{subequations}
The evaluation of equation(32b) is similar to the evaluation in Section IV that gave equation (26) differing only in the use of the following
\begin{equation}\label{33}
\begin{split}
\int d^{3}p \vert\vec{p}\vert^{2} \exp {(-\beta c (\vec{p}^{2} + m^{2}c^{2})^{\frac{1}{2}}} & = \frac{4\pi}{c^{2}} \dfrac{\partial ^{2}}{\partial \beta^{2}} \left[ (\frac{m^{2}c}{\beta}) K_{2}(mc^{2}\beta) \right]\\
& \quad -m^{2} c^{2}\left[ 4\pi (\frac{m^{2} c}{\beta} K_{2}(mc^{2} \beta) \right] .
\end{split}
\end{equation}
Equation (33) suggests that the $ \frac{1}{c^{2}} $ may not be consistent because positive powers of $ c^{2} $ may just creep in. Fortunately, as I show in Appendix B, there is no such danger. And just like in Section IV, after a long and tedious computation, the resulting relativistic correction to $ Z_{hy} $ to order $ \frac{1}{c^{2}} $ is 
\begin{equation}\label{34}
\Delta Z_{hy} = \left[ \frac{18N}{\pi m \beta} Z_{hy} - \frac{18aN}{4m\beta}\left( \frac{2\pi}{\beta \lambda N} \right)^{\frac{3}{2}N-1} V^{\frac{2}{3}-N} \exp {(-\frac{1}{8} \beta \lambda N V^{\frac{2}{3}} )} Z_{fr} \right] ,
\end{equation}
where $ Z_{fr} $ is the partition function for N free relativistic particles given by equations (28) and (29).
Notice that equation (34) mirrors equation (26) the difference being instead of $ Z_{nr} $ and $ Z_{fnr}$ we have instead $ Z_{hy} $ and $ Z_{fr} $. It should be expected that the thermodynamics of the relativistic system should be more accurately computed using the hybrid expansion.
\section{\label{sec:level6}Conclusion}
In this paper, I show how to compute the relativistic partition function up to $ \frac{1}{c^{2}} $  by making use of the relativistic potential, which I derived in a previous paper. To illustrate the method, I applied the formalism to a system of N particles with two-body interaction given by a harmonic potential. I then computed the partition function in two ways - (1) with a non-relativistic Hamiltonian with both kinetic energy term and two-body potential non-relativistic (see sections III and IV ), (2) with a hybrid expansion, with the kinetic term fully relativistic because its contribution to the partition function is exactly known in a closed form and relativistic potential expanded in powers of $ \frac{1}{c^{2}} $ (see Section V and Appendix B). The paper solved the problem of the relativistic correction to thermodynamics up to order $\frac{1}{c^{2}} $.

I end with a discussion on satisfying the Poincare algebra in relativistic particle-particle dynamics. Inherent in relativistic particle dynamics is the time-delay in the interaction. In my previous paper, I showed that it is this time-delay that prevents the generators of space-time transformations from satisfying the Poincare algebra. However, as I noted in the introduction, there are previous works that claim the contrary. But their Hamiltonian and other  generators are different from what I derived. I think this is an issue that still needs clarification.   

\begin{acknowledgments}
I would like to thank Ms Antonietta Villaflor of the College of Science Library of the University of the Philippines for helping me get a number of papers that I used in this work. I would also like to thank Felicia Magpantay for correcting my Latex file.
\end{acknowledgments}

\begin{appendix}

\renewcommand{\theequation}{A-\arabic{equation}}
\setcounter{equation}{0}  
\setcounter{section}{0}
\section*{Appendix A. The Value of $ \Lambda \Xi $}
\setcounter{section}{0}
 
 Here, I prove equation (24) of the paper that says $ \Lambda \Xi = -\frac{\pi}{40} $. To show the result, first I do an integration by parts in $ \Xi $ to get
 \begin{equation}\label{A.1}
 \Xi = \lim_{v_{0} \to  0} \frac{1}{4} \left( \frac{1}{v_{0}} \right)^{4} - \frac{1}{20} \int_{0}^{\infty} dv \frac{1}{v^{4}} \sin v.
 \end{equation}
 The product of the two factors then become
 \begin{equation}\label{A.2}
 \Lambda \Xi = \Lambda  \lim_{v_{0} \to 0} \frac{1}{4} \left( \frac{1}{v_{0}} \right)^{4} - \frac{1}{20} \left( \int_{0}^{\infty} du u^{3} \sin u \right) \left( \int_{0}^{\infty} dv \frac{1}{v^{4}} \sin v \right).
 \end{equation}
 I evaluate the second term by transforming to plane polar coordinates by defining
 \begin{subequations}\label{A.3}
 \begin{gather}
 u = r\cos \theta, \label{first}\\
 v = r\sin \theta, 
 \end{gather}
 \end{subequations}
The integral in the second term of equation (A.2) becomes
\begin{equation}\label{A.4}
int = \int_{0}^{\frac{\pi}{2}} \dfrac{(\cos \theta)^{3}}{(\sin \theta )^{4}}(\frac{1}{2})\int_{-\infty}^{\infty} dr \sin (r\cos \theta) \sin (r \sin \theta),
\end{equation}
which after a few steps give $ int = \frac{\pi}{2} $. This makes the second term of equation (A.2) equal to the answer $ -\frac{\pi}{40} $. 

Now I show that the first term of equation (A.2) vanishes. For this, I make use of another representation for $ \Lambda $ from \cite{Gradshteyn} page 420, formula 3.761.2,
\begin{equation}\label{A.5}
\Lambda = \lim_{u_{0} \to 0} \frac{i}{2} \left[ \exp {(-i2\pi)} \Gamma(4,iu_{0}) -  \exp {(+i2\pi)} \Gamma(4,-iu_{0}) \right] ,
\end{equation}
where $ \Gamma(4,x) $ is an incomplete gamma function. This has a series expansion given by  \cite{Gradshteyn} page 941 formula 8.354.2
\begin{equation}\label{A.6}
\Gamma(\mu,x) = \Gamma(\mu) - \sum_{n=0}^{\infty} (-1)^{n} \dfrac{x^{(\mu+n)}}{n!(\mu + n)!}.
\end{equation}
This results in power series that begins with
\begin{equation}\label{A.7}
\Lambda =\lim_{u_{0} \to 0} \frac{i}{2} \left[ 2i \frac{(u_{0})^{5}}{5!} + ...\right] .
\end{equation}
This makes the first term of equation(A.2) equal to
\begin{equation}\label{A.8}
first term = \lim_{u_{0} \to 0} \lim_{v_{0} \to 0} (\frac{1}{v_{0}^{4}}) \frac{i}{2} \left[ 2i\frac{(u_{0})^{5}}{5!} + ...\right].
\end{equation}
Since the $ u_{0} \rightarrow 0 $ term goes to zero faster than the $ v_{0} \rightarrow 0 $ term diverges, this term is zero. Thus, it has been shown
\begin{equation}\label{A.9}
 \Lambda \Xi = - \frac{\pi}{40}.
\end{equation}

\renewcommand{\theequation}{B-\arabic{equation}}
\setcounter{equation}{0}
\setcounter{section}{0}  
\section*{Appendix B. Consistency of hybrid expansion}
The hybrid expansion takes a full relativistic treatment of the kinetic term but expands the relativistic two-body potential term in powers of $ \frac{1}{c^{2}} $. As argued in Section V, the momentum integrals in the correction just may pull down positive powers of $ c^{2} $ when computing . Here I show that this does not happen. 

The momentum integrals in the $ \frac{1}{c^{2}} $ corrections in the two-body potential yields the partition function of relativistic kinetic term $ Z_{rkin} $ given by equation (29) plus the spatial integral factors, which has no c dependence, and finally an extra factor given by
\begin{equation}\label{B.1}
m^2c^{2}z\frac{1}{K_{2}(z)}\dfrac{\partial^{2}}{\partial z^{2}}\left[ \frac{1}{z} K_{2}(z) - m^{2}c^{2}\right],
\end{equation}
where $ z = mc^{2}\beta $. For the $ \frac{1}{c^{2}} $ expansion to be consistent, this term better be of the order of $ (\frac{1}{c^2})^{0} $ and higher, i.e., $ (\frac{1}{c^{2}})^{n} $, with n =1, 2,.... Equation (B.1) simplifies to
\begin{equation}\label{B.2}
m^{2}c^{2}\left[ \frac{12}{z^{2}} + \frac{5}{z} \dfrac{K_{1}(z)}{K_{2}(z)} + \dfrac{K_{0}(z)}{K_{2}(z)}\right] - m^2 {c^{2}}.
\end{equation}
From \cite{Gradshteyn} page 970 formula 8.486.17, which says
\begin{equation}\label{B.3}
K_{2}(z) = \frac{2}{z} K_{1}(z) + K_{0}(z),
\end{equation}
the problematic $ m^{2}c^{2} $ term cancels out leaving the term

\begin{equation}\label{B.4}
m^{2}c^{2}\left[ \frac{12}{z^{2}} + \frac{3}{z} \dfrac{K_{1}(z)}{K_{2}(z)} \right] .
\end{equation}
The first term of above is order $ \frac{1}{c^{2}} $, which means it can be neglected because as equation (31) shows, there is already a $ \frac{1}{c^{2}} $ factor that goes with $ \Delta Z_{hy}$. The second term is $ \frac{3m}{\beta} \dfrac{K_{1}(z)}{K_{2}(z)} $. From \cite{Weisstein}
\begin{equation}\label{B.5}
K_{n}(z) = \sqrt{\frac{\pi}{2z}} \dfrac{\exp {-z}}{(n-\frac{1}{2})!} \sum_{r=0}^{\infty} \dfrac{(n-\frac{1}{2})!}{r!(n-r-\frac{1}{2})!} (2z)^{-r} \Gamma(n+r+\frac{1}{2}).
\end{equation}
From this follows
\begin{equation}
\dfrac{K_{1}(z)}{K_{2}(z)} = 1 + O(\frac{1}{c^{2}}) + O(\frac{1}{c^{4}}) + ....
\end{equation}
Equation (B.4) gives $ \frac{3m}{\beta} $ resulting in $ \Delta Z_{hy} $ given by equation (34).

\end{appendix}

\end{document}